\pdfoutput=1
\documentclass[reprint,linenumbers,aip, graphicx]{revtex4-1}
\usepackage{graphicx}
\usepackage{dcolumn}
\usepackage{bm}
\usepackage{bm}
\usepackage{hyperref}
\hypersetup{
   colorlinks,=true,
    citecolor=blue,
    filecolor=blue,
    linkcolor= blue,
   urlcolor=blue
}
\begin{document}


\title{Anisotropic charge transport in non-polar GaN quantum wells (QWs): polarization-induced line charge and interface roughness scattering} 



\author{Aniruddha Konar}
\email[]{akonar@nd.edu}
\author{Tian Fang}
\author{Nan Sun}
\author{ Debdeep Jena}
\affiliation{ Department of Physics and Department of Electrical Engineering, University of Notre Dame, Notre Dame, USA 46556}


\date{\today}

\begin{abstract}
 Charge transport in GaN quantum well (QW) devices grown in the non-polar   direction is theoretically investigated . Emergence of a new form of anisotropic line charge scattering mechanism originating from anisotropic rough surface morphology in conjunction with in-plane built-in polarization is proposed. It is shown that such scattering leads to a large anisotropy in carrier transport properties, which is partially reduced by strong isotropic optical phonon scattering.
\end{abstract}

\maketitle 
The growth of non-polar {\it{m}}-plane (1$\bar{1}$00) and {\it{a}}-plane (11$\bar{2}$0) GaN has attracted a lot of interest recently \cite{WaltereitNature00,BenjaminJEM05}. Though the built-in polarization field in traditional polar-GaN ($c$-plane) has been exploited to achieve dopant free high electron mobility transistors (HEMTs), for optical devices, polarization field plays a negative role due to quantum confined Stark effect (QCSE). Moreover, c-plane GaN-based enhancement-mode (E-mode) HEMTs have very low threshold voltage (V$_{th}\sim$ 1 V) due to inherent presence of polarization induced electron gas (2DEG). High threshold voltage (V$_{th}\sim$ 3 V) is required for high-voltage switching operations.  So the recent trend is to explore optical and transport properties of GaN grown in the non-polar direction. Recently E-mode transistor with high threshold voltage ($\ge$ 2 V) has been achieved \cite{FujiwaraAPE09} with non-polar GaN heterojunction. \\A characteristic feature of non-polar GaN surfaces grown directly on foreign substrates is extended stripe or slate like morphology perpendicular to the $c$ axis \cite{WangAPL04, ChenAPL02, Paskova05, BenjaminJEM05, HiraiAPL07}, regardless of growth methods. The origin of this rough surface morphology has been attributed to i) replication of substrate morphology \cite{SunJAP02}, ii) extended basal-plane stacking faults (BSFs) \cite{BenjaminJEM05, HiraiAPL07}, and iii) anisotropic diffusion barrier of Ga adatoms \cite{BrandtPhysica04, LymperakisPRB09}. This striated morphology has been conjectured qualitatively to be responsible for experimentally observed conductivity anisotropy for bulk GaN films \cite{McLaurinJAP06} as well as thin GaN-QWs \cite{BaikIEEE10}. No microscopic theory is available in existing literature for a quantitative estimation of this electrical anisotropy. This paper develops  a microscopic theory of anisotropic carrier transport in non-polar GaN QWs with a quantitative estimation of the transport anisotropy.\\
\begin{figure}[htp]%
\includegraphics*[width=85 mm]{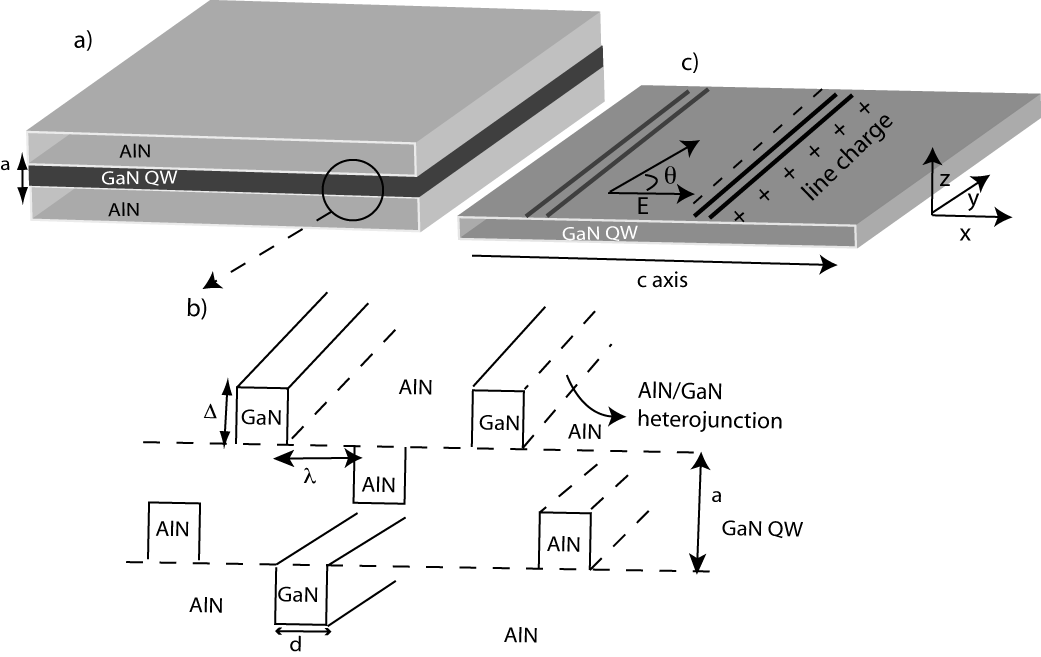}
\caption{%
  Non-polar GaN quantum well: a)QW of width $a$ sandwiched between AlN barriers, b) schematic diagram of interface roughness at GaN/AlN interface, c) polarization induced line charge at each step edge of roughness. }
\label{Fig1}
\end{figure}
Let us consider a thin non-polar GaN QW of thickness $a$ sandwiched between two aluminum nitride (AlN) layers as shown in Fig.\ref{Fig1}a). A common source of disorder for non-polar GaN QW is striped surface morphology as mentioned in the previous section. We model this rough surface morphology as a local variation of QW thickness ( see Fig.1) The variation of QW thickness alone causes local shifts of the conduction band edge, resulting in carrier scattering commonly known as interface roughness scattering (IR) in literature. The interface roughness (IR) can be modeled by local thickness fluctuations $\Delta(x)$ of the non-polar GaN QW with a spatial correlation $\langle\Delta(x)\Delta(0)\rangle=\Delta^{2}\exp[-|x|/\sqrt{2}\Lambda]$, where  $\Delta$ is the average height of roughness and $\Lambda$ is the in-plane correlation length between two roughness steps. Denoting  the envelope function of conduction electrons in the n$^{th}$ subband of GaN QWs as $|n,r,k\rangle=\sqrt{2/a}\sin(\pi nz/a/)e^{i{\vec{k}}{\cdot}{\vec{r}}}/\sqrt{L_{x}L_{y}}$, the square of the unscreened intra-subband IR matrix element of scattering from initial momentum state  ${\bf{k_{i}}}(k_{i},\theta)$ to a final momentum state  ${\bf{k_{f}}}(k_{f},\theta^{\prime})$ in the m$^{th}$ subband can be written as 
\begin{eqnarray}
|v_{IR}(q_{x})|^2 &=&\frac{m^2}{L_{x}}\Big(\frac{\pi^{2}\hbar^2\Delta}{m^{\star}a^3}\Big)^2\frac{2\sqrt{2}\Lambda}{2+(q_{x}\Lambda)^2}\delta_{q_{y},0},
\label{Eq1}
\end{eqnarray}
where, ${\bf{q}}={\bf{k_{f}}}-{\bf{k_{i}}}$ ,$ L_{x}$ and $L_{y}$ are the macroscopic lengths of the QW in $x$ and $y$ directions, $\delta_{(..)}$ is  Kronecker delta function and $\hbar$ is the reduced Planck constant. The problem of IR scattering is well known and has been investigated for two and one dimensional electron gases in many semiconductors, including nitrides.  What is new in non-polar structure is the polarization induced bound charges associated with each interface roughness step. The thickness modulation of  QW leads to GaN/AlN heterojunctions at each roughness center as depicted in Fig.\ref {Fig1} b). The difference of in-plane polarization of GaN and AlN creates line-charge dipoles by inducing fixed charges at opposite faces of each step (as shown in Fig.\ref{Fig1}.b). Assuming each rough step is infinitely extended along the $y$ direction (see ref. 6-9), the electrostatic potential at any point {\textbf{r}} $(x,y,z)$ in the QW arising from a polarization line-charge dipole of a roughness step at a point ($x_{i},0$) of height $\Delta (x_{i})$ and lateral width $d$ is given by
\begin{eqnarray}
V(x,y,z)&=&v_{+}(x,y,z)-v_{-}(x,y,z)\nonumber \\
&=&\frac{e\lambda_{\pi}}{4\pi\epsilon_{0}\kappa}\Big[ln\frac{(x-x_{i}-d/2)^{2}+z^{2}}{(x-x_{i}+d/2)^{2}+z^{2}}\Big],
\label{Eq2}
\end{eqnarray}
where $e$ is the electron charge, $\epsilon_{0}$ is the free-space permittivity, $\kappa$ is the relative dielectric constant of GaN and $\lambda_{\pi}(x_{i})=\Big|P_{GaN}-P_{AlN}\Big|\Delta(x_{i})/e$ is the effective line charge density. Note that, the potential is independent of $y$ due the symmetry of the problem.  The scattering matrix element of transition from state $|n,r,k_{i}\rangle$ to state $|m,r,k_{f}\rangle$ can be written as $v(x_{i},q)=\langle m,r,k_{f}|v(x,y,z)|n,r,k_{i}\rangle$, where
\begin{eqnarray}
v(x_{i},q)&=&\Big(\frac{e\lambda_{\pi}}{\epsilon_{0}\kappa|q_{x}d|}\Big)e^{-iq_{x}x}\sinh\Big(\frac{|q_{x}d|}{2}\Big)F_{nm}(q_{x}a)\nonumber \\
&&\times\delta_{q_{y},0}.
\label{Eq3}
\end{eqnarray}
 In the above equation, $F_{nm}(q_{x}a)$ is the form factor arising from the quasi-2D nature of the electron gas. For our choice of the envelope function, the form factor can be calculated analytically. It approaches unity  ($F_{nm}(q_{x}a)\rightarrow 1)$, both for long wavelength ($q\rightarrow 0$) as well as for for a very thin QW ($a\rightarrow 0$).\\
\begin{figure}
\includegraphics [width=80 mm]{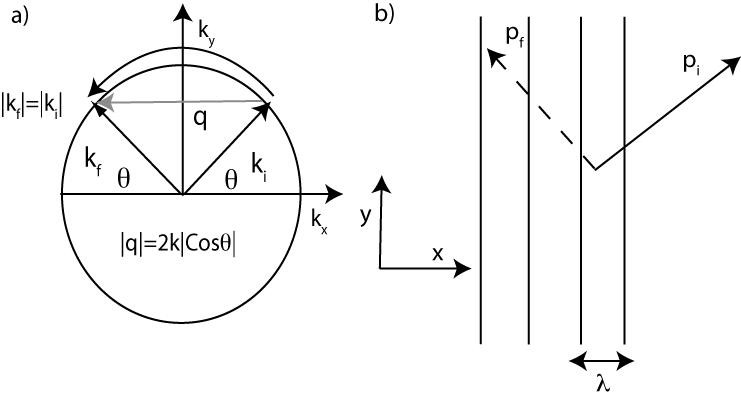}
\caption{schematic diagram of momentum change in anisotropic elastic scattering.
  a):momentum space. Clearly transverse wave vector ($k_{y}$) of carriers parallel to line charges is conserved while the net momentum change is $|2k\cos\theta|$, b) in real space electron's movement is shown.  }
\label{Fig2_1}
\end{figure}
Eq. (3) represents the Fourier-transformed electrostatic potential from a line charge dipole associated with a single roughness step. As two steps are correlated, the dipole potential arising from them are also correlated. If there are N numbers of roughness steps, the square of the matrix element of the dipole potential summed over all scatterers is given by
\begin{equation}
|v_{dip}(q_{x})|^2=v(0,q)^{2}L_{x}n_{dip}\Big[1+\frac{2\sqrt{2}n_{dip}\Lambda}{2+(q_{x}\Lambda)^2}\Big],
\label{Eq4}
\end{equation}
where, $n_{dip}=N/L_{x}$ is the average roughness density (/cm) at the interface.The other two important scattering mechanisms are: i) remote charge impurity (RI) scattering, and ii) polar optical phonon scattering. To have free carriers, the QW is doped remotely. If $t$ is the distance between the QW and the remotely doped layer, the unscreened matrix element of scattering can be written as \cite{BookDavis}
\begin{equation}
|v_{RI}(q)|^2=\Big(\frac{e^2F_{mn}(qa)}{2\epsilon_{0}\kappa }\Big)^2\frac{e^{-2qt}}{q^2}.
\end{equation}
Polar optical phonon (POP) scattering rates under Davydov-Shmushkevich  scheme ($\hbar\omega_{0}\gg k_{B}T$), where phonon emission is assumed to be instantaneous,  has been analytically calculated by Gelmont et. al \cite{GelmontJAP95}. The only difference between Gelmont's approach and our calculation is the form factor, which was calculated by Gelmont et. al using a Fang-Howard type wavefunction, whereas, in our case, it is calculated using infinite well (hard wall boundary conditions) -type wavefunction.\\
Among the scattering potentials considered above, the polarization-induced line charge potential and interface-roughness potential are anisotropic in nature. This leads to anisotropic scattering events which is captured by the Kronecker delta function ($\delta_{q_{y},0}$) appearing in the corresponding matrix elements of scattering. Fig.\ref{Fig2_1} shows a typical anisotropic scattering event where, electron's momentum along $y$ direction remains unchanged while the momentum along $x$ direction is reversed. For such anisotropic scattering, net momentum change is $|q|\equiv |q_{x}|=2k\cos\theta$, where $\theta$ is the angle of the incoming wave vector with $x$ axis. This is in striking difference with RI scattering, where,  $|q|=2k\cos\psi$, where $\psi=\theta^{\prime}-\theta$; the angle between initial and final wave vector of scattering (see inset of Fig.\ref{Fig2}b). 
For line charge scattering and IRs, scattering strengths are equally important at all angles except at $\theta=\pm\pi/2$ where scattering time diverges (no scattering), whereas for RI scattering only small angle scatterings rates dominate. For such anisotropic scatterers, an angle averaged single relaxation time approximation (RTA) formalism fails, and one needs to look for either variational \cite{AndoJPSJ03} or numerical solutions of Boltzmann transport equation (BTE).  Recently, Schliemann and Loss \cite{LossPRB03}(SL) have proposed an exact solution of BTE in the presence of anisotropic scatterers which we use in this work.\\
\begin{figure}[t]%
\includegraphics[width=85mm] {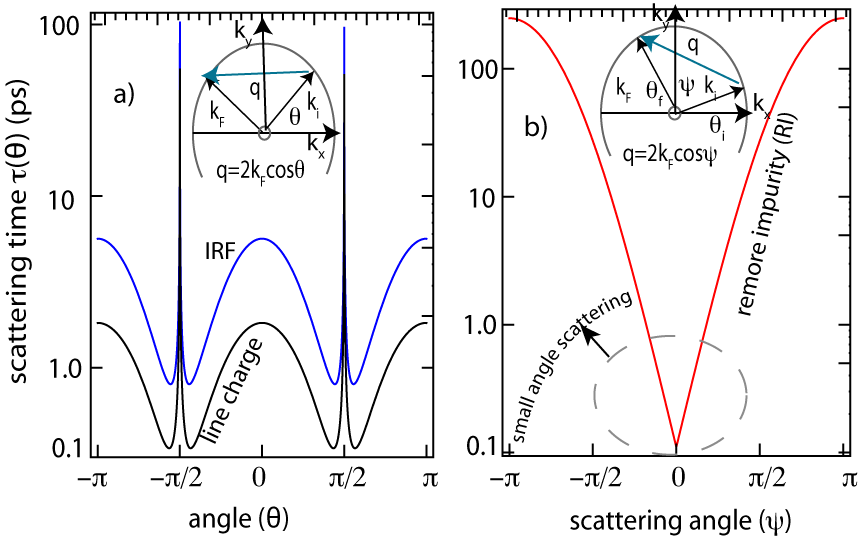}
\caption{%
  a): angle ($\theta$) dependent scattering time for line-charge (black) and IR (blue) scattering. b): RI scattering time as a function of angle ($\psi$) between initial and final wave vector of scattering.}
\label{Fig2}
\end{figure}
To investigate the effect of anisotropic scattering events on experimentally measurable transport quantities (such as electron mobility, conductivity etc), it is sufficient to consider carrier transport along the two principal directions $x$ and $y$. Transport coefficients in any arbitrary direction, in principle, can be  obtained by coordinate transformation. We first consider charge transport along $x$. Under the application of a small electric field ${\vec{E}}=(E,0)$ in the QW-plane, the non-equilibrium part of carrier distribution function under the SL scheme can be written as
\begin{equation}
g_{k}^{\parallel}(\theta)=-e\Big(-\frac{\partial f_{0}}{\partial\mathcal{E}_{k}}\Big)A_{k}^{\parallel}(\theta){\bf{v(k)}}. {\bf{E}},
\end{equation}
 where $f_{0}$ is the equilibrium Fermi-Dirac distribution function, ${\bf{v(k)}}$ is the group velocity, and $\mathcal{E}_{k}$ is the kinetic energy of an electron in the GaN QW. The coefficient $A_{k}^{\parallel}(\theta)$ is defined as
 \begin{equation}
  A_{k}^{\parallel}(\theta)=\frac{\tau^{\parallel}({\bf{k}})}{1+\Big(\frac{\tau^{\parallel}({\bf{k}})}{\tau^{\perp}({\bf{k}})}\Big)^2},
  \end{equation}
 where $\tau^{\parallel} (\tau^{\perp})$ is the scattering time parallel (perpendicular) to applied field calculated using Fermi's golden rule. Defining the current density as ${\vec{J}}=2e/(L_{x}L_{y})\sum_{k}g_{k}{\vec{v_{k}}}$, the expression for mobility for a degenerate electron gas along the $x$ direction can be calculated as
 \begin{equation}
  \mu_{xx}=\frac{e}{\pi m_{xx}^{*}}\int_{0}^{2\pi}d\theta A^{\parallel}(k_{f},\theta)\times\cos^{2}\theta
  \end{equation}
where $m^{*}_{xx}$ is the electron effective mass in GaN along the $c$ axis. A similar expression can be derived for $\mu_{yy}$. It can be seen that for isotropic scattering, $A_{k}^{\parallel}=A_{k}^{\perp}=\tau(k_{f})$; which leads to conventional isotropic mobility  $\mu_{xx}=\mu_{yy}=e\tau/m^{*}$. In general, the integral appearing in Eq. (9) is evaluated numerically for the complicated angular dependence of the coefficient $A_{k}^{\parallel}(\theta)$. Nevertheless, analytical expressions of mobilities can be evaluated under certain approximation. For a very thin QW, $F_{mn}(qa)=1$, and electron mobility for anisotropic scatterer can be written as
\begin{eqnarray}
\mu_{xx}^{dip}&=&\frac{e\hbar\epsilon_{0}\kappa a^{*}_{B}}{m^{*}_{xx}n_{dip}(\lambda_{\pi}d)^2}\sqrt{\frac{8n_{s}}{\pi^{3/2}}}\mathcal{I}_{1}\Big(\frac{q_{TF}}{2k_{F}}\Big) \ \ \ \ \ \ \mbox{and},\nonumber \\  
\mu_{xx}^{IR}&=& \frac{ea^6}{4\hbar\Delta^2\Lambda}\sqrt{\frac{n_{s}}{\pi^{9/2}}}\mathcal{I}_{2}\Big(\frac{q_{TF}}{2k_{F}}\Big).
\end{eqnarray}
Where, $q_{TF}$ is the 2D Thomas-Fermi wave vector \cite{BookDavis}, $n_{s}$ is the equilibrium carrier density in the QW, $k_{F}$ is the Fermi wave vector, and $a^{*}_{B}$ is the effective Bohr radius \cite{BookDavis}.  The dimensionless integrals $\mathcal{I}_{1}\Big(\frac{q_{TF}}{2k_{F}}\Big)$ and $\mathcal{I}_{2}\Big(\frac{q_{TF}}{2k_{F}}\Big)$ can be evaluated exactly \cite{integral}.\\
For numerical calculations, a nominal set of parameters are used to describe the GaN QW roughness:  $(\Delta,d)=(3.19, 5.2)$\AA, whereas other parameters such as carrier density $(n_{s})$, temperature $(T)$ and the QW thickness $(a)$ are varied within an experimentally relevant range. An isotropic effective mass $m^{*}_{xx}=m^{*}_{yy}=0.23m_{0}$ ($m_{0}$ is the mass of a bare electron) is assumed, since the effective mass along the $c$ axis ($m^{*}_{c\parallel}=0.228m_{0}$) and perpendicular to the $c$ axis  ($m^{*}_{c\perp}=0.237m_{0}$) has negligible anisotropy \cite{JAPMeyer03} . An important parameter describing the magnitude of roughness is the in-plane correlation length $\Lambda$. In Fig.\ref{Fig3}, we plot the IR scattering rates with in-plane correlation length $\Lambda$ for different Fermi energies. The scattering rate exhibits a peak for $\Lambda\approx 1/k_{x}$, where IR scattering matrix element is maximized, and then decays slowly with $\Lambda$ on either sides of the peak. For a particular scattering rate, it is possible to find two values of $\Lambda$, on each side of the peak. Since $\Lambda$ is a two-valued function of scattering rates (mobility), it is necessary \cite{SakakiAPL} to determine $\Lambda$ from temperature dependent mobility data.  Due to unavailability of transport data for non-polar GaN at present, we assume $\Lambda$=1.5 nm - a typical value used \cite{YuAPL07} for polar GaN HEMTs. It should be noted that the line-charge scattering rates will have similar but weaker dependence on $\Lambda$ due to inclusion of correlated dipole scattering in Eq. 4.\\
 \begin{figure}[t]
\includegraphics[width=85mm] {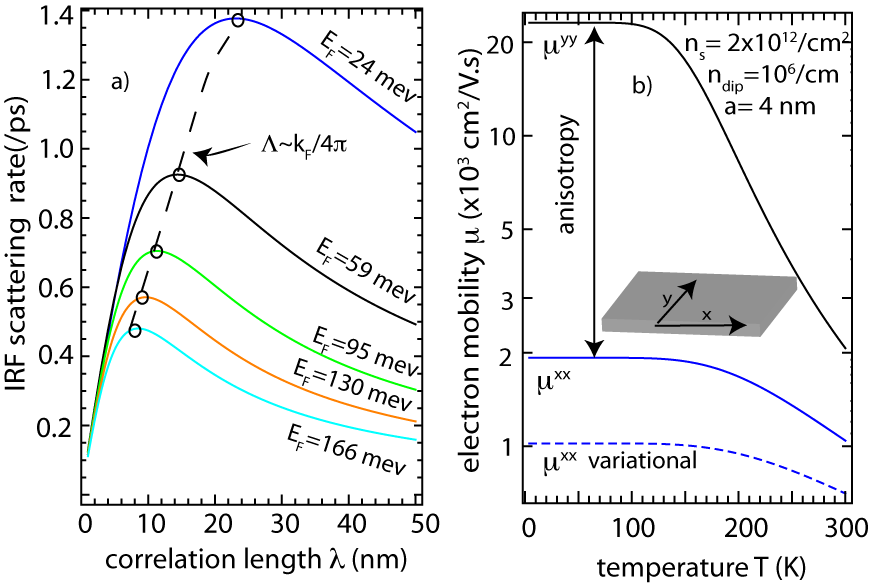}
\caption{ a): IRF scattering vs in-plane correlation length $\Lambda$ for various carrier densities (Fermi energies), b) electron mobility along two principal directions x (blue) and y (black) as a function of temperatures. The dashed blue line is the longitudinal mobility using variational technique.}
\label{Fig3}
\end{figure}
The fact that anisotropic scatterers do not hinder electrons moving along $y$ direction, in principle, should result in a higher mobility along the $y$ direction. Fig.\ref{Fig3}b) captures this anisotropy. The temperature dependent electron mobility is evaluated along two  principal directions $x$ and $y$. At low temperatures, $\mu_{yy}$, limited by RI scattering, is significantly higher than longitudinal mobility $\mu_{xx}$ which is limited by anisotropic IR and line charge scattering in addition to RI scattering. At room temperature, isotropic polar optical phonon scattering tends to reduce the anisotropy in mobility by equally affecting $\mu_{xx}$ and $\mu_{yy}$. We have assumed distance of remote doping layer $t$= 3 nm \cite{}. Larger values of t will increase the mobility anisotropy even more at low temperatures by increasing $\mu_{yy}$ exponentially. For example, a value of $t= 10$ nm would result in $\mu_{yy}\approx 10^5$ cm$^2$/V.s \cite{BookJena} for $n_{s}=10^{12}$/cm$^2$ at low temperatures; approximately 3 times higher mobility compared to $t$= 3 nm case (see Fig.\ref{Fig3}b). The dashed blue curve in Fig. 3b) shows the mobility ($\mu_{xx}$) calculated using the variational principle \cite{AndoJPSJ03, BookAM}. The difference of numerical values of mobility between variational and SL scheme stems out from the fact that variational techniques gives the lower bound of mobility.\\
Both line-charge and IR scattering matrix elements are strong decreasing functions of electron's kinetic energy, which implicitly depends on the magnitude of the momentum transfer $q_{x}$ in scattering processes. For a  degenerate electron concentration the mobility is effectively determined the carriers at Fermi level. Hence $q_{x}\approx 2k_{f}=\sqrt{2\pi n_{s}}$, where $n_{s}$ is the equilibrium carrier density.
 As a result, both IR and line charge scattering rates decrease, which in effect, reduces mobility anisotropy  with increasing carrier density $n_{s}$  as illustrated in Fig.\ref{Fig4}a). Similarly, for wide QWs, IR ($\mu_{IR}\sim a^{6}$) and line charge scattering are unimportant and polar optical phonon scattering is the dominant scattering mechanism. Consequently, $\mu_{xx}$ and $\mu_{yy}$ tend to approach to the same limiting value  (determined by the POP scattering rate), and mobility anisotropy is completely washed out (see Fig.\ref{Fig4}b)) for $a> 8 $ nm. At this point, we want to stress upon the fact that in our numerical calculations, we have used a minimal set of parameters for interface roughness ($\Delta,d \sim $ monolayer). In practice, values of these parameters differ from sample to sample and experimentally extracted set of parameters should be used for a more accurate quantitative description of transport anisotropy.  
 \begin{figure}[t]
\includegraphics[width=85mm] {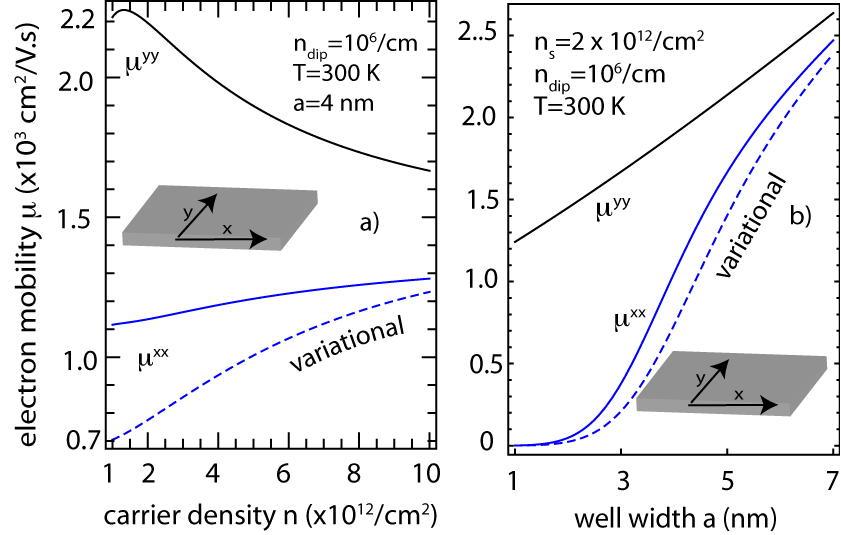}
\caption{electron mobility along two principal axes x (blue) and y (black)--- a): as a function of carrier densities and b): as a function of well width. Note that mobility anisotropy decreases with increasing carrier density or well width. Dashed blue curves represent longitudinal mobility calculated using variational technique.}
\label{Fig4}
\end{figure}
Two effects have not been taken account in our model - i) scattering from charged BSFs and ii) anisotropic strain at GaN/AlN interface. While scattering from charged BSFs is expected to enhance transport anisotropy in addition to the anisotropy presented here, strain-induced piezoelectric polarization will alter transport anisotropy by altering bound line charge density at each roughness step. How to incorporate these two effects in our model remains an open problem and should be addressed in future for a more complete and accurate description of charge transport in non-polar GaN QWs.\\
In summary, a theory of charge transport in non-polar GaN QWs has been presented. We show that extended defects, together with the in-plane polarization of non-polar GaN-based devices act as anisotropic scattering centers. At low temperatures, the mobility shows highly anisotropic behavior for thin QWs. At room temperature, the magnitude of transport anisotropy is reduced by strong isotropic polar optical phonons scattering. It is shown that variational technique overestimates the transport anisotropy compared to exact solution of BTE.\\
The authors would like to acknowledge  J. Verma, S. Ganguly (University of Notre Dame) for helpful discussions and National Science Foundation (NSF), Midwest Institute for Nanoelectronics Discovery (MIND) for the financial support for this work.

\end{document}